\documentclass{elsarticle}
\usepackage{graphicx}

\begin{document}
\title{Coupled mode theory for acoustic resonators}
\author[IF]{Dmitrii N. Maksimov\corref{FN1}\fnref{FN2}}
\cortext[FN1]{Corresponding author}
\fntext[FN2]{Tel.: +7 391 2494538, Fax: +7 391 2438923}
\ead{mdn@tnp.krasn.ru}
\author[IF]{Almas F. Sadreev}
\author[SFU]{Alina A. Lyapina}
\author[SFU]{Artem S. Pilipchuk}
\address[IF]{ LV Kirensky Institute of Physics, 660036, Krasnoyarsk,
Russia}
\address[SFU]{ Siberian Federal University, 660080, Krasnoyarsk,
Russia}
\date{\today}
\begin{abstract}

We develop the effective non-Hermitian Hamiltonian
approach for open systems with Neumann boundary conditions.
The approach can be used for calculating the scattering matrix and the scattering function
in open resonator-waveguide systems. In higher than one dimensions the method represents acoustic
coupled mode theory in which the scattering solution within an open resonator is found in the
form of expansion over the eigenmodes of the closed resonator decoupled from the waveguides.
The problem of finding the transmission spectra is reduced to solving a set of linear equations
with a non-Hermitian matrix whose anti-Hermitian term accounts for coupling between the resonator eigenmodes and the scattering
channels of the waveguides. Numerical applications to acoustic two-, and three-dimensional resonator-waveguide
problems are considered.
\end{abstract}

\maketitle


\section{Introduction}
The approach of the effective non-Hermitian Hamiltonian \cite{physrep,Dittes,Ingrid}
have found numerous applications in various branches of physics including
atomic nuclei \cite{Weidenmuller,Sokolov}, chaotic billiards
\cite{Stockmann, Alhassid, Pichugin, Stockmann1, Akguc, Auerbach}, tight-binding models
\cite{Datta, SR,Sadreev,Hatano,Hatano1}, potential scattering \cite{Savin}, photonic crystals \cite{Bulgakov}, etc.
The  objective  of  the  present  paper  is to revisit the concept of the effective non-Hermitian Hamiltonian
in application to open resonators with the Neumann boundary conditions. The problem of resonant scattering typically
 involves a resonator
(which could be an atom, atomic nucleus, quantum dot, microwave
or acoustic cavity  {\it etc}) and one, two or more scattering channels that couple
the resonator to the environment. The mainstream idea
 is to split the full Hilbert space into
subspaces: subspace $B$ formed by the eigenfunctions of discrete spectrum localized
within the scattering center, and subspace $C$ which spans the extended eigenfunctions of the scattering channels.
Therefore, the exact description of open system meets a problem of matching the wave functions of discrete
and continuous spectra. In 1957 Livshits \cite{Livshits} and independently Feshbach in 1958
\cite{Feshbach} introduced the idea to project the total Hilbert space onto the discrete states of subspace
$B$. Given the Hamilton operator of the whole system as
\begin{equation}\label{Htot}
\mathcal{\widehat{H}}=\mathcal{\widehat{H}}_B+\sum_C(\mathcal{\widehat{H}}_C+V_{BC}+V_{CB})
\end{equation}
the projection onto the discrete subspace leads to the concept of the effective non-Hermitian
Hamiltonian \cite{physrep, Dittes, Ingrid, Feshbach}
\begin{equation}
\mathcal{\widehat{H}}_{eff} = \mathcal{\widehat{H}}_B + \sum_C V_{BC} \frac{1}{E^+ - \mathcal{\widehat{H}}_C} V_{CB}.
\label{Heff}
\end{equation}
Here $\mathcal{\widehat{H}}_B$ is the Hamiltonian of the closed system, $\mathcal{\widehat{H}}_C$ is the Hamiltonian of the scattering channel $C$,
 $V_{BC},~V_{CB}$ stand for the coupling matrix elements between the
eigenstates of $\mathcal{\widehat{H}}_B$ and the eigenstates of the scattering channel $C$, and $E$ is the energy of scattered
particle (wave). The term $E^{+} =E+i0$ ensures that only outgoing waves
will be present in the solution after the scattering occurs. As a result the effective
Hamiltonian (\ref{Heff}) is a non-Hermitian matrix with complex eigenvalues $z_{\lambda}$
which determine the energies and lifetimes of the resonant states as
$Re(z_{\lambda})$, and $-2Im(z_{\lambda})$ correspondingly \cite{physrep,Ingrid}.
If the propagation band of waveguide is infinite then the effective non-Hermitian Hamiltonian takes the most simple form
widely used in the scattering theory \cite{physrep, Dittes, Weidenmuller, Sokolov, Auerbach}
\begin{equation}\label{Heffphen}
    \mathcal{\widehat{H}}_{eff}=\mathcal{\widehat{H}}_B-i\pi\sum\limits_{C=1} W_CW_C^{\dagger},
\end{equation}
where $W_C$ is a column matrix whose elements account for the coupling of each individual inner state to the scattering
channel $C$, and the symbol $\dagger$ stands for Hermitian transpose.  The scattering matrix $\mathcal{S}_{CC'}$ is then
given by the inverse of $E-\mathcal{\widehat{H}}_{eff}$
 \cite{Dittes,Stockmann}
\begin{equation}
\mathcal{S}_{C'C}(E)=\delta_{C'C}-2\pi i W_{C'}^{\dagger}\frac{1}{E-\mathcal{\widehat{H}}_{eff}}W_C,
\end{equation}
where $\delta_{C'C}$ is the Kronecker delta.

The approach of the effective non-Hermitian Hamiltonian for two dimensional resonator-waveguide systems controlled by the
Schro\"odinger equation was previously addressed
in Refs. \cite{Stockmann,Pichugin, Stockmann1}. In particular in Ref. \cite{Pichugin} the authors derived exact formulas for the coupling
matrices $W_C$ for both Dirichlet and Neumann Boundary conditions on the boundary of the resonator.
What is more interesting, however,
it was shown \cite{Pichugin} that only in the case of the Neumann boundary condition the approach is stable with respect
to truncation of the discrete basis to a finite number of eigenstates thanks to the absolute convergence of the spectral
sum for the reaction matrix. In the above references \cite{Stockmann, Pichugin, Stockmann1} the resonator-waveguide problem was
considered in the context of quantum scattering. This imposes restrictions on the applicability of the effective
non-Hermitian Hamiltonian because one normally requires the Dirichlet boundary conditions on the infinitely high
ward-wall boundary. Although, there are techniques to improve the convergence of the spectral sum \cite{Lee, Schanz}
in the Dirichlet case, in general one would resort to the mixed boundary conditions
applying the Dirichlet boundary conditions on the physical boundaries while the Neumann boundary condition
is applied on the waveguide-resonator interface \cite{Pichugin}. One the other hand, the Neumann boundary conditions are the boundary
conditions for the pressure field on a sound hard boundary. That prompts us to apply the effective
non-Hermitian Hamiltonian formalism to acoustic scattering problem. In essence,
the proposed approach relying on the spectral properties of a closed resonator decoupled from the environment is analogous
to the coupled mode theory \cite{Haus, Suh} which is a very popular tool for analyzing resonant scattering in optics.
Technically, the optical coupled mode theory \cite{Haus, Suh} represents a method for finding transmission spectra
from a set of linear equations with a matrix analogous to Eq.(\ref{Heffphen}) in which the diagonal Hermitian term consists
of the eigenfrequencies of the resonator, while the second ani-Hermitian term accounts for the coupling between the eigenmodes
of the resonator with the scattering channels.
In this paper we will focus on developing acoustic coupled mode theory including applications to discretized systems
which render the method applicable to
open acoustic resonators of arbitrary shape in which the eigenfunctions could not be found analytically.

The article is organized as follows. In Sec. \ref{sec2} we consider a simple one-dimensional tight-binding model
which is aimed to illustrate our approach to derive the effective non-Hermitian Hamiltonian. In Sec. \ref{sec3}
we extend our results to 2D case and demonstrate the connection between the discrete model based on the finite-difference
representation of the Helmholtz equation and the continuous model based on the eigenfunctions of the partial differential
equation. In Sec. \ref{sec3} we demonstrate an application of the acoustic coupled mode theory for finding the transmission
spectra and scattering functions in a realistic 3D structure. Finally, we conclude in Sec. \ref{sec5}.


\section{The effective non-Hermitian Hamiltonian for one-dimensional
system with the Neumann boundary conditions} \label{sec2}

\begin{figure}[h]
\includegraphics[height=6cm,width=10cm,clip=]{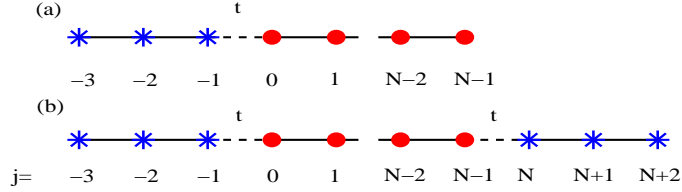}
\caption{(Color online) One-dimensional tight-binding chain (resonator) of N sites (red circles)
coupled to one (a) or two (b) semi-infinite waveguides (blue stars). The coupling between the resonator
and the waveguides (dashed lines) is controlled by hopping matrix element $t$. All other sites are coupled via
unit hopping matrix element (solid lines).} \label{fig1}
\end{figure}

There are many approaches to establish the effective non-Hermitian  Hamiltonian formalism
\cite{physrep,Dittes,Ingrid, Stockmann,Datta,SR,Savin}. In this paper we adopt a variation of the method recently developed
for two-particle Bose-Hubbard lattice model \cite{Maksimov}. To describe the method we begin with the simplest
possible one-dimensional tight-binding model
which consists of one-dimensional resonator coupled to one or two half-infinite wires (waveguides). The systems under
consideration are sketched in Fig. \ref{fig1}
We will see later that this approach can be easily generalized to higher dimensions as well as applied to  the continuous limit which
corresponds to 2D and 3D acoustic problems.

\subsection{Tight-binding chain coupled to a single waveguide}

For the model depicted in Fig. \ref{fig1} (a) the 1D stationary Schr\"{o}dinger equation
takes the following form
\begin{equation}\label{1dtb}
\mathcal{\widehat{H}}\psi_j=-t_j^{L}\psi_{j-1}-t_j^{R}\psi_{j+1}+(2-\delta^j_{N-1})\psi_j=E\psi_j; \ j=-\infty, \ldots, N
\end{equation}
where
\begin{equation}
t_j^{(L)}= \left\{\begin{array}{cc}
t & \mbox{if $j=0$}; \\ 1 & \mbox{otherwise},
\end{array}\right.
\end{equation}
while
\begin{equation}
t_j^{(R)}= \left\{\begin{array}{cc}
t & \mbox{if $j=-1$};
\\ 0 & \mbox{if $j=N-1$}
\\ 1 & \mbox{otherwise}.
\end{array}\right.
\end{equation}
One can see that the parameter $t$ in this model  controls the coupling between the resonator and the waveguide.
Meanwhile, because the Neumann BC at site $N$ in the discrete case are written as
\begin{equation}\label{Neumann}
\psi_{N}-\psi_{N-1}=0,
\end{equation}
the term $\delta^j_{N-1}$ is introduced at the end of the tight-binding chain to account for the Neumann boundary condition.

The key idea of our approach is to split the total space into two subspaces.
The first subspace corresponds to
the resonator with $j=0, 1, \ldots, N-1$
while the second subspace corresponds to the waveguide with $j< 0$. Respectively we
present the solution of the Eq. (\ref{1dtb}) in the following form
\begin{equation}\label{solution}
\psi(j)=\left\{\begin{array}{cc}
\psi_B(j) & \mbox{if $j=0, 1, \ldots, N-1$}; \\ a_L^{(+)}\psi_{+}(j)+a_L^{(-)}\psi_{-}(j)
& \mbox{$j\leq 0$},
\end{array}\right.
\end{equation}
where $\psi_B(j)$ is the solution within the resonator
\begin{equation}\label{psiB}
\psi_B(j)=\sum_{m=1}^N\chi_m\psi_m(j),
\end{equation}
with  $\psi_m(j)$ as the eigenfunctions of the resonator. Notice, that here $\chi_m$ are unknown coefficients
which are yet to be found for Eq.(\ref{solution}) to be the solution of the discrete Schr\"{o}dinger equation
(\ref{1dtb}). The eigenfunctions $\psi_m(j)$ are given by the following equation.
\begin{equation}\label{boxeig}
\psi_1(j)=\sqrt{\frac{1}{N}}, ~ \psi_m(j)=\sqrt{\frac{2}{N}}\cos(k_m(j+1/2)), ~
k_m=\frac{\pi (m-1)}{N}, m=2,3,\ldots,N.
\end{equation}
with the corresponding eigenenergies given by
\begin{equation}
E_m=2-2\cos(k_m).
\end{equation}
We note in passing that $\psi_m(j)$ Eq. (\ref{boxeig}) are the eigenvectors of the $N\times N$ matrix Hamiltonian operator
with the Neumann boundary conditions (\ref{Neumann}) at $j=0$ and $j=N-1$
\begin{equation}
\label{HB}
\mathcal{\widehat{H}}_N=\left(\begin{array}{ccccccc} 1 & -1 & 0 & \cdots & 0 & 0 & 0\cr
                       -1 & 2 & -1& \cdots & 0 & 0 & 0\cr
                        0 & -1& 2 & \cdots & 0 & 0 & 0\cr
                        \vdots&\vdots&\vdots&\ddots&\vdots&\vdots&\vdots\cr
                        0 & 0 & 0 & \cdots & 2 & -1 & 0\cr
                        0 & 0 & 0 & \cdots & -1 & 2 & -1\cr
                        0 & 0 & 0 & \cdots & 0 & -1 & 1
\end{array}\right).
\end{equation}
The solution in the waveguide is written as a superposition of ingoing and outgoing waves
\begin{equation}
\label{waves}
\psi_{(\pm)}(j)=\frac{1}{\sqrt{4\pi\sin(k)}}e^{\pm ikj},
\end{equation}
with $a_L^{(\pm)}$ as the incoming and outgoing amplitudes.
The waves $\psi_{\pm}(j)$ are eigenfunctions of the tight-binding
infinite wire $-\psi_{(\pm)}(j-1)-\psi_{(\pm)}(j+1)+2\psi_{(\pm)}(j)=E\psi_{(\pm)}(j)$ with the dispersion relation
\begin{equation}\label{disp}
E=2-2\cos(k).
\end{equation}
Obviously, in this present model the $\mathcal{S}$-matrix
contains only reflection coefficient $\mathcal{S}=r$ with $|r|=1$, i.e.,
only the phase of the wave changes in the scattering.

In order to find the equation for coefficients $\chi_m$, we evaluate the scalar product
$\langle \psi_m|\mathcal{\widehat{H}}-E|\psi\rangle$=0 with operator $\mathcal{\widehat{H}}$
implicitly defined through Eq. (\ref{1dtb}). We mention in passing that the bra-ket notation will be adopted throughout
the paper for vector quantities to comply with
the previous works \cite{Dittes,Sadreev,Savin}. Specifically, $|\ldots\rangle$ stands for a column vector,
while $\langle\ldots|={|\ldots\rangle}^{\dagger}$ is the
corresponding Hermitian conjugate row vector.
\begin{eqnarray}\label{projbox}
&\langle \psi_m|\mathcal{\widehat{H}}-E|\psi\rangle=(2-E)\chi_m-\sum\limits_{j=0}^{N-2}\psi_m(j)\psi(j+1)&\nonumber\\
&-\sum\limits_{j=1}^{N-1}\psi_m(j)\psi(j-1)-\psi_m(N-1)\psi(N-1)-t\psi_m(0)\psi(-1)&\nonumber\\
&=(2-E)\chi_m-t\psi_m(0)\psi(-1)&\nonumber\\
&-\sum\limits_{n=1}^N\chi_n\left[\sum\limits_{j=0}^{N-2}\psi_m(j)\psi_n(j+1)+
\sum\limits_{j=1}^{N-1}\psi_m(j)\psi_n(j-1)-\psi_m(N-1)\psi_n(N-1)\right]&\nonumber\\
  &=-E\chi_m+\sum\limits_{n=1}^N\chi_n\langle\psi_m|\mathcal{\widehat{H}}_N|\psi_n\rangle+\sum\limits_{n=1}^N\chi_n\psi_m(0)\psi_n(0)
-t\psi_m(0)\psi(-1)&\nonumber\\
&=(E_m-E)\chi_m+\sum\limits_{n=1}^{N}\chi_n\psi_m(0)\psi_n(0)-\frac{t}{\sqrt{4\pi\sin k}}\psi_m(0)
\left[a_L^{(+)}e^{-ik}+a_L^{(-)}e^{ik}\right].&
\end{eqnarray}
On the other hand the scalar quantity $\langle \psi_{(-)}|\mathcal{\widehat{H}}-E|\psi\rangle=0$
could be evaluated as
\begin{eqnarray}\label{projW}
& \langle \psi_{(-)}|\mathcal{\widehat{H}}-E|\psi\rangle=
-\psi_{(-)}(-1)\left[\psi(-2)+t\psi_B(0)-(2-E)\psi(-1)\right]&\nonumber\\
& -\sum\limits^{-2}_{j=-\infty}\psi_{(-)}(j)
\left[\psi(j-1)+\psi(j+1)-(2-E)\psi(j)\right] &\nonumber\\
&=\frac{1}{4\pi\sin k}\left[a_L^{(+)}+a_L^{(-)}\right]e^{-ik}-
\frac{t}{\sqrt{4\pi\sin k}}e^{-ik}\sum\limits_{n=1}^{N}\chi_n\psi_n(1).&
\end{eqnarray}
The resulting expressions could be presented in a matrix form as
\begin{equation}\label{matrix}
\left\{
\begin{array}{cc} E_m-E +W_LW_L^{\dagger}& -\frac{t}{\sqrt{4\pi\sin k}}e^{ik}W_L
\\ -\frac{t}{\sqrt{4\pi\sin k}}e^{-ik}W_L^{\dagger} & \frac{1}{4\pi\sin k}e^{-ik}
\end{array}
\right\} \left\{
\begin{array}{c}
|\chi\rangle \\  a^{(-)}_L
\end{array}
\right\}= \left\{
\begin{array}{c}
\frac{t}{\sqrt{4\pi\sin k}}e^{-ik}W_L a_L^{(+)}\\  -\frac{1}{4\pi\sin k}e^{-ik}a_L^{(+)}
\end{array}
\right\},
\end{equation}
where $|\chi\rangle$ is the column vector of coefficients $\chi_m$ which define the scattering function $\psi_B(j)$ through
Eq. (\ref{psiB}), while $W_L$ is a column vector with its elements given by
\begin{equation}
\label{W}
\{W_L\}_m=\psi_m(1).
\end{equation}
From the second row of Eq. (\ref{matrix}) we have
$$a_L^{(-)}=-a_L^{(+)}+t\sqrt{4\pi\sin(k)}W_L^{\dagger}|\chi\rangle.$$
Substituting this in the first row of Eq. (\ref{matrix}) we obtain
\begin{equation}\label{LS}
(\mathcal{\widehat{H}}_{eff}-E)|\chi\rangle=-it\sqrt{\frac{1}{\pi}\sin(k)}W_La_L^{(+)},
\end{equation}
where operator $\mathcal{\widehat{H}}_{eff}$ could be easily recognized as the effective non-Hermitian
Hamiltonian
\begin{equation}
    \mathcal{\widehat{H}}_{eff}=E_m\delta_{mn}+(1-t^2e^{ik})W_LW_L^{\dagger}.
\end{equation}

\subsection{Transmission through one dimensional system}

The model considered in the previous subsection was introduced with the only goal to illustrate
our approach to derive the effective non-Hermitian Hamiltonian.
Let us now consider a less trivial example, namely, the tight-binding model depicted in Fig. {\ref{fig1}} (b).
Repeating the calculations from the previous subsection one can easily find that the effective non-Hermitian Hamiltonian now takes the
following form
\begin{equation}\label{Heffm2}
    \mathcal{\widehat{H}}_{eff}=E_m\delta_{mn}+(1-t^2e^{ik})\sum_{C=L,R}W_CW_C^{\dagger}.
\end{equation}
with
\begin{equation}
\{W_R\}_m=\psi_m(N-1).
\end{equation}
The equation for the scattering function now reads
\begin{equation}\label{LS2}
(\mathcal{\widehat{H}}_{eff}-E)|\chi\rangle=-it\sqrt{\frac{1}{\pi}\sin(k)}\sum_{C=L,R}W_Ca^{(+)}_C,
\end{equation}
while the reflection amplitudes can be found as
\begin{equation}\label{ac}
a_C^{(-)}=-a_C^{(+)}+\sqrt{4\pi\sin(k)}{W}_C^{\dagger}|\chi\rangle
\end{equation}
The $\mathcal{S}$-matrix is implicitly defined through the following equation connecting incoming and
outgoing amplitudes
\begin{equation}\label{S}
\left(\begin{array}{c} a_{L}^{(-)}\cr a_{R}^{(-)}\end{array}\right)=\left(\begin{array}{cc}
r & t'\cr t& r'\end{array}\right)\left(\begin{array}{c} a_{L}^{(+)}\cr a_{R}^{(+)}\end{array}\right)
\end{equation}
where $a_{L}^{(-)}, a_{R}^{(-)}$ are the right and left outgoing amplitudes,
$r, r'$ are the reflection coefficients,
and $t, t'$ are the transmission coefficients.
Combining the above equation we find the $\mathcal{S}$-matrix in the following form \cite{Dittes,SR}
\begin{equation}\label{S-matrix1d}
\mathcal{S}_{C,C'}=-\delta_{C,C'}-it\sqrt{2\sin (k)}{W}_{C}^{\dagger}
\frac{1}{\mathcal{\widehat{H}}_{eff}-E}t\sqrt{2\sin (k)}{W}_{C'}
\end{equation}

After some algebra it could be easily found that in the coordinate representation $\mathcal{\widehat{H}}_{eff}$
takes the following form
\begin{equation}\label{Heffsite}
\langle j|\mathcal{\widehat{H}}_{eff}|j'\rangle=\mathcal{\widehat{H}}_N+(1-t^2e^{ik})(\delta_{j,N-1}\delta_{j',N-1}+
    \delta_{j,0}\delta_{j',0})=\mathcal{\widehat{H}}_D-t^2e^{ik}(\delta_{j,N-1}\delta_{j',N-1}+
    \delta_{j,0}\delta_{j',0})
\end{equation}
where $\mathcal{\widehat{H}}_D$ is the tight-binding Hamiltonian of the chain of length $N$ with the Dirichlet
boundary conditions
\cite{Datta,SR}
\begin{equation}
\label{HD}
\mathcal{\widehat{H}}_D=\left(\begin{array}{ccccccc} 2 & -1 & 0 & \cdots & 0 & 0 & 0\cr
                       -1 & 2 & -1& \cdots & 0 & 0 & 0\cr
                        0 & -1& 2 & \cdots & 0 & 0 & 0\cr
                        \vdots&\vdots&\vdots&\ddots&\vdots&\vdots&\vdots\cr
                        0 & 0 & 0 & \cdots & 2 & -1 & 0\cr
                        0 & 0 & 0 & \cdots & -1 & 2 & -1\cr
                        0 & 0 & 0 & \cdots & 0 & -1 & 2
\end{array}\right)
\end{equation}
It is also worth mentioning that the effective non-Hermitian Hamiltonian in the Dirichlet representation has the following form \cite{SR}
\begin{equation}\label{Heffm}
    \mathcal{\widehat{H}}_{eff}=E_m\delta_{mn}-t^2e^{ik}\sum_{C=L,R}{\overline{W}}_C\overline{W}_C^{\dagger}
\end{equation}
where $\{\overline{W}_L\}_m=\overline{\psi}_m(0)$, $\{\overline{W}_R\}_m=\overline{\psi}_m(N-1)$ with $\overline{\psi}_m(j)$ as the eigenfunctions
of the Dirichlet eigenvalue proplem. One can see that Eq.(\ref{Heffsite}) is fully consistent with the earlier findings \cite{Datta,SR}.
However the present approach is much simpler, in particular it is free from evaluation of principal value integrals \cite{Datta,SR}.

\begin{figure}[ht]
\includegraphics[height=8cm,width=8cm,clip=]{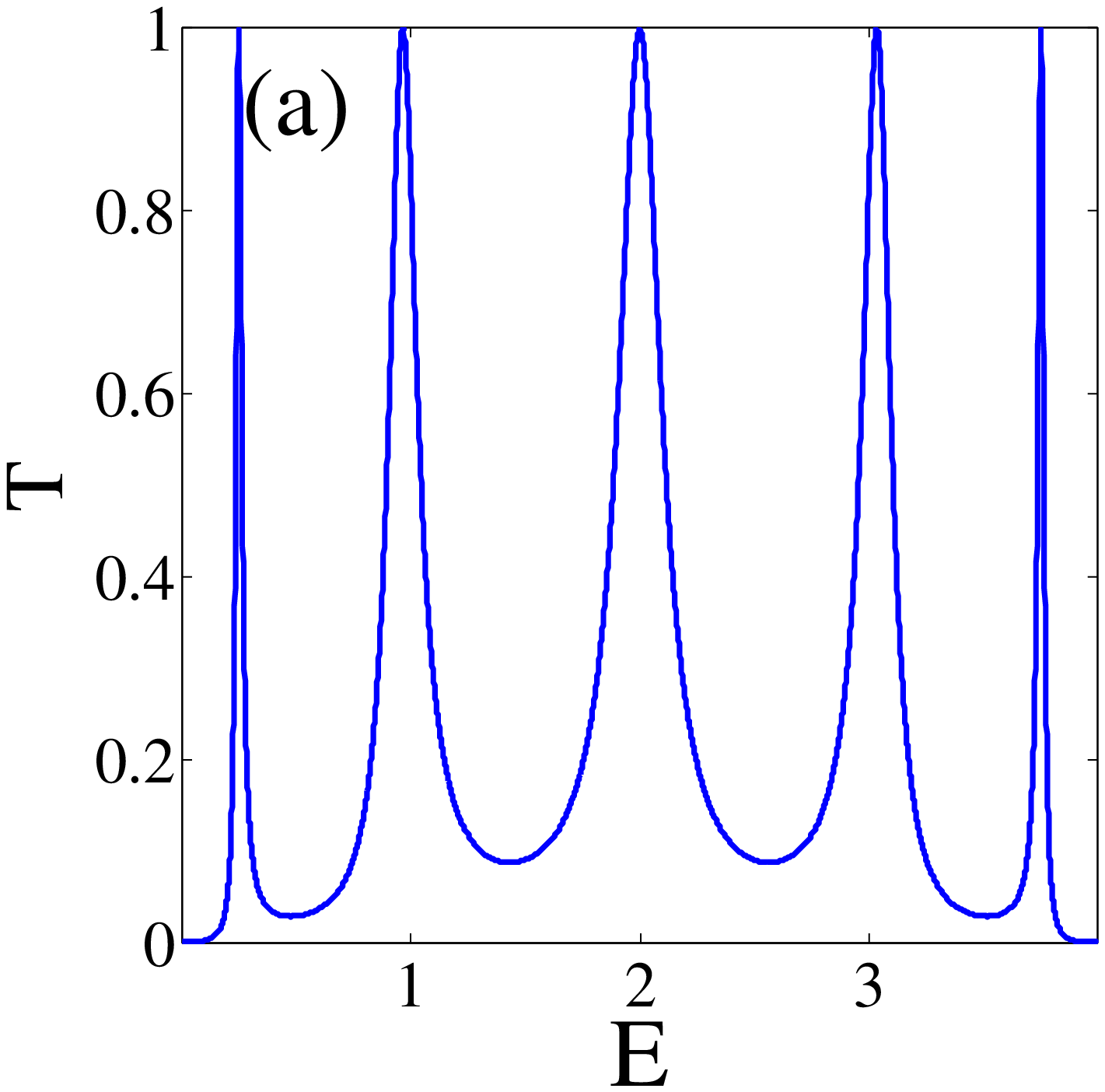}
\includegraphics[height=8cm,width=8cm,clip=]{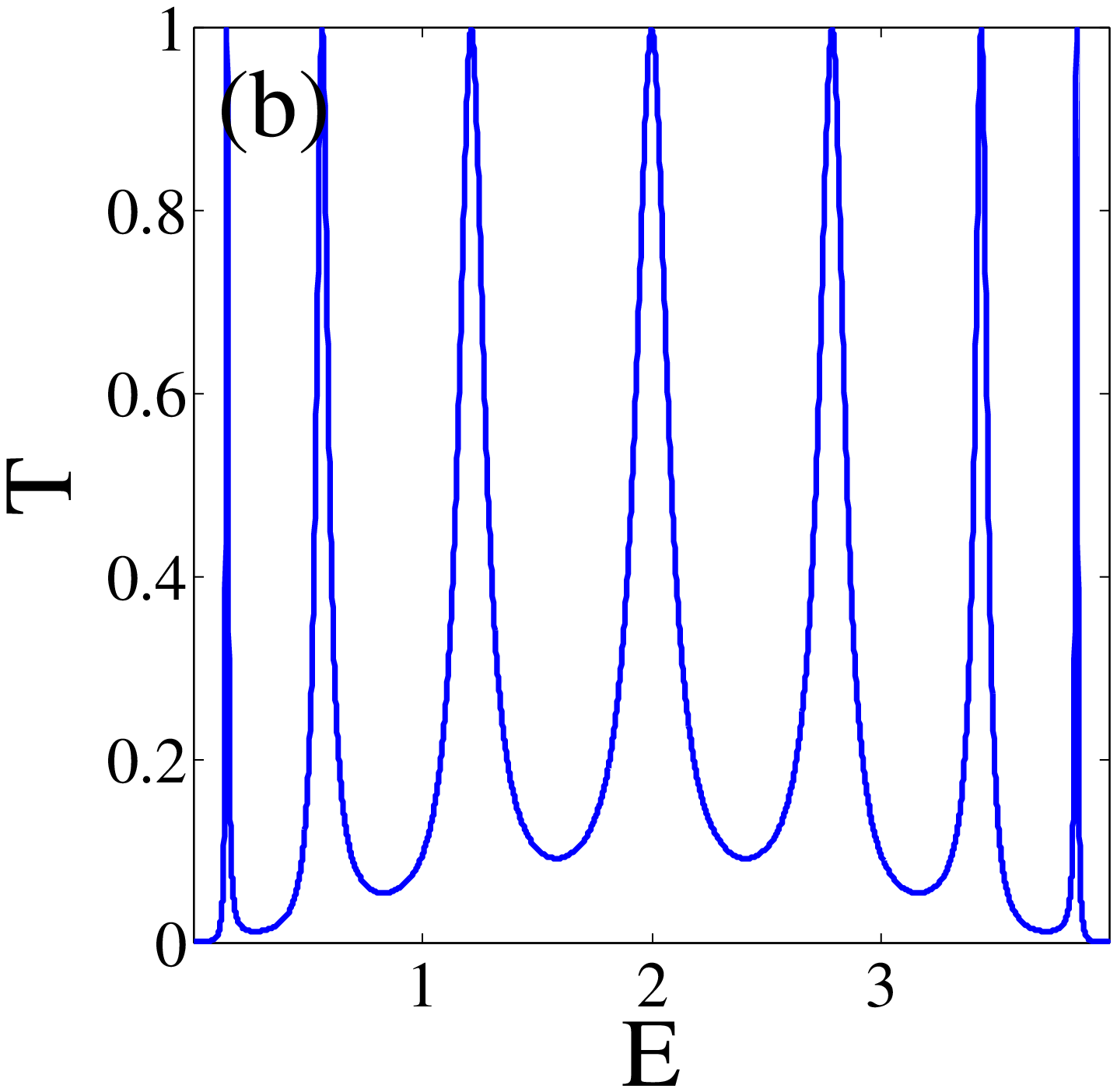}
\caption{(Color online)Transmission spectra of tight-binding chains of 5 (a) and 7 (b) sites, $t=0.4$}\label{fig2}
\end{figure}

Finally, to illustrate the effective non-Hermitian Hamiltonian approach we plot transmission probability $T$ vs. energy $E$
in Fig. \ref{fig2}. To obtain the transmission probability $T$ one can directly apply Eq. (\ref{S-matrix1d}).
However, this approach would be numerically inefficient because it involves evaluation of the inverse of the matrix $\mathcal{\widehat{H}}_{eff}-E$.
In practice to speed-up the performance one would first use Eq. (\ref{LS2}) to find the interior wave function $|\chi\rangle$ and then
use Eq. (\ref{ac}) for finding the reflection amplitudes.
The data in Fig. \ref{fig2} obviously coincide with the transmission probability through tight-binding chain with the Dirichlet BC according to \cite{SR}.
The reason for this is that the resonator-waveguide boundary is artificial and, thus, one has freedom in choosing the boundary conditions \cite{Savin}
as long as the resulting equations (\ref{Heffm2}) is consistent with the total Hamiltonian of the system (\ref{1dtb}).
However, as it will be seen below, the transmission coefficient depends on the choice of the boundary conditions at physical boundaries of the resonator.


\section{Two-dimensional model} \label{sec3}

\begin{figure}[ht]
\includegraphics[height=7cm,width=11cm,clip=]{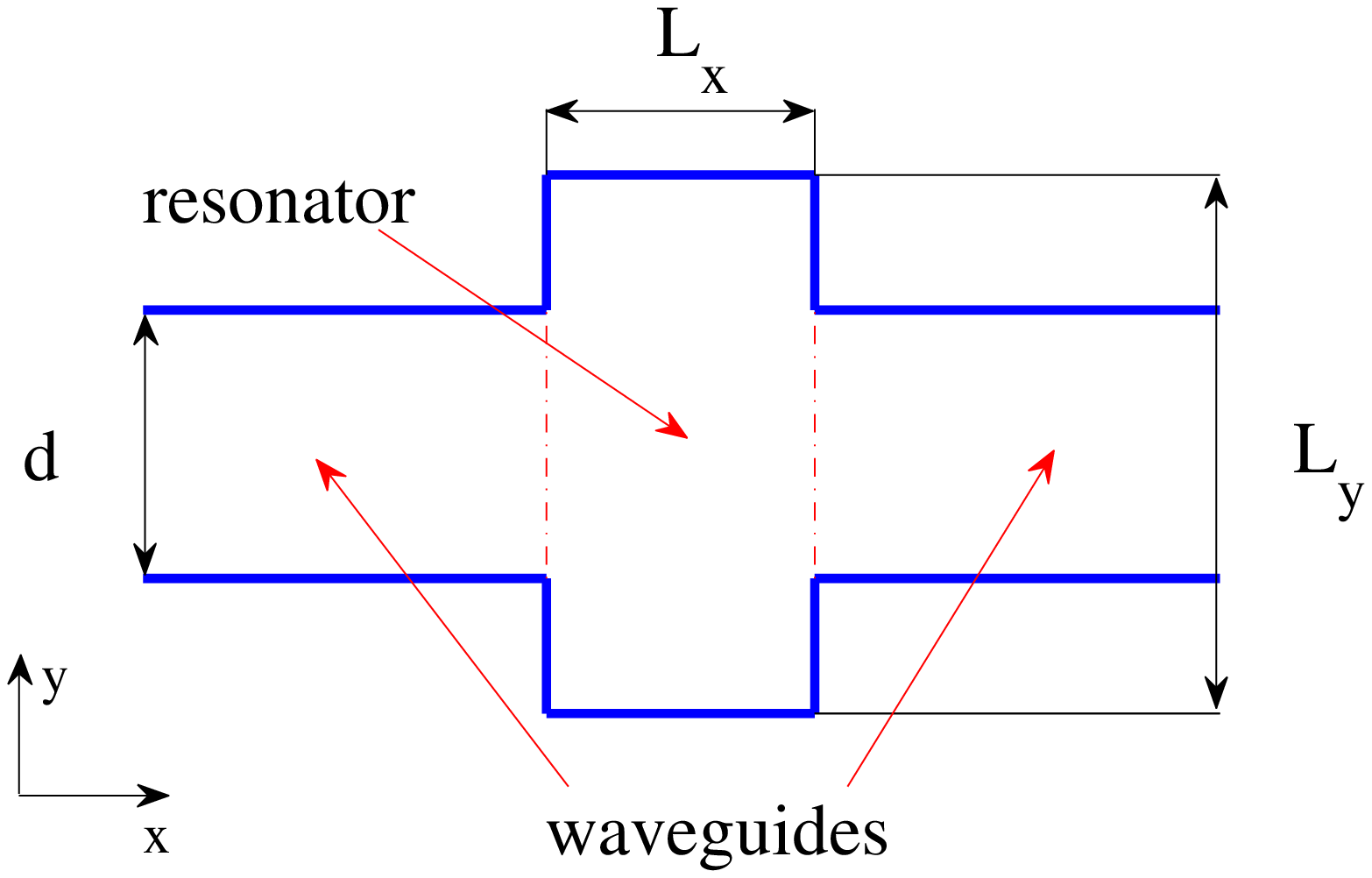}
\caption{(Color online) 2D acoustic resonator coupled two waveguides.}\label{fig3}
\end{figure}

Before proceeding to the effective non-Hermitian Hamiltonian we would like to make some remarks to
draw an analogue between the tight-binding model and acoustics. The stationary acoustic field
satisfies the Helmholtz equation
\begin{equation}\label{Helmholtz}
    \nabla^2\psi+\frac{\omega^2}{c^2}\psi=0,
\end{equation}
where $\psi$ is the pressure, $c$ - the speed of sound, and $\omega$ is the frequency.
In what follows $c$ is set equal to unity. The condition on the sound hard boundary is the Neumann boundary
condition which requires the normal derivative of the pressure field to be equal to zero on the boundary $S$
\begin{equation}\label{NBC}
{\left.\frac{\partial \psi}{\partial n}\right|}_{S}=0
\end{equation}
In this section we consider the resonator shown in Fig. \ref{fig3}.
This system can be seen as a limiting case of three-dimensional planar duct-cavity-duct acoustic system with a rectangular cross-section
if the thickness of the resonator along the $z$-axis is substantially smaller than the length and the width in the $x$ and $y$-dimensions.

To find the numerical solution the Helmholtz equation (\ref{Helmholtz}) could be written through the centered second
difference scheme as
\begin{equation}\label{tbSE}
\frac{1}{a_0^2}\left(\psi_{l+1,j}+\psi_{l-1,j}+\psi_{l,j+1}+\psi_{l,j-1}-4\psi_{l,j}\right)+\frac{\omega^2}{c^2}\psi_{l,j}=0,
\end{equation}
where $a_0$ is the step-size in space.
The resulting expression  (\ref{tbSE}) could be viewed as an eigenvalue problem
$$\widehat{\mathcal{H}}\psi=E\psi, $$
for the tight-binding "Hamiltonian" $\widehat{\mathcal{H}}$ with four off-diagonals which carry the nearest neighbor hopping
matrix element $1/a_0^2$ , while the "energy" is given by $$E=\omega^2/c^2.$$
Notice that Eq. (\ref{tbSE}) preserves all symmetries of the system in Fig. \ref{fig3} in the discretized coordinates $x_j=a_0j, y_l=a_0l$.
therefore one can easily introduce the discretized Neumann boundary conditions (\ref{Neumann}) as
$$ \psi_{\pm N_l,j}-\psi_{\pm (N_l-1),j}=0, \ \  \psi_{l, \pm N_l}-\psi_{l,\pm (N_l-1)}=0$$
on the vertical and horizontal boundaries correspondingly (see Fig. \ref{fig3}) with $N_l$ and $N_j$ being the indexes of the edge sites. At this point we would like
to notice that our choice of the system
and the discretization method are solely aimed at simplifying the ensuing algebra.
If the configuration of the system is more involved
then the finite-difference scheme would not be the method of choice. We will see, however, that in the continuous limit $a_0\rightarrow 0$ the resulting
expressions are not dependant on the orientation of the grid. On the other hand if a more sophisticated meshing procedure, like hp-adaptive
finite-element method \cite{Demkowicz}, is applied one would use the same technique to derive the effective non-Hermitian Hamiltonian under the only conditions that
the grid is locally regular on each waveguide-resonator interface and preserves the translational symmetry in the waveguides.

\subsection{The effective non-Hermitian Hamiltonian for 2D lattice model} \label{discretesub}

The technique described in the previous section can be easily generalized for 2D case.
Let $P$ be the number of sites across the waveguide and $N$ the number of sites along the resonator, then the channel functions are given by
\begin{equation}\label{wavedisc1}
\psi_{L,p}^{(\pm)}(j,l)=\phi_p(l)\phi_L^{(\pm)}(j), \ j<0
\end{equation}
and
\begin{equation}\label{wavedisc2}
\psi_{R,p}^{(\pm)}(j,l)=\phi_p(l)\phi_R^{(\pm)}(j), \ j>N-1,
\end{equation}
where $\phi_q(l)$ is the transversal wavefunction of the waveguide
\begin{equation}
\phi_1(l)=\sqrt{\frac{1}{P}}, \ \phi_p(l)=\sqrt{\frac{2}{P}}\cos\left(\frac{\pi (p-1) (l+P/2)}{P} \right), \ p=2, \ 3, \ \ldots P,
\end{equation}
while the travelling waves $\phi_R^{(\pm)}(j)$ and $\phi_L^{(\pm)}(j)$ are
\begin{equation}\label{trw}
\phi_R^{(\pm)}(j)=\frac{a_0}{\sqrt{4\pi\sin(k_pa_0)}}\exp{(\pm ik_pa_0j )}, \ \
\phi_L^{(\pm)}(j)=\frac{a_0}{\sqrt{4\pi\sin(k_pa_0)}}\exp{(\pm ik_pa_0j \mp ik_pa_0N)}.
\end{equation}
The corresponding dispersion relationship is given by
\begin{equation}\label{disp22}
E_p=\frac{1}{a_0^2}\left[4-2\cos(a_0k_p)-2\cos\left(\frac{\pi(p-1)}{P}\right)\right].
\end{equation}
Using (\ref{disp22}) one can easily check that the waves (\ref{trw}) obey the normalization condition
\begin{equation}\label{norma}
   \langle j|j' \rangle= \int\limits_{E_{min}}^{E_{max}} dE \psi_p^{(\pm)}(j)^{*}\psi_p^{(\pm)}(j')=
    \frac{1}{2\pi}\int_{-\pi}^{\pi}d\alpha e^{i\alpha(j-j')}=\delta_{jj'}.
\end{equation}

The scattering domain is a rectangular box with $N$ sites in the direction parallel to the transport
axis and $M$ sites in the transversal direction. Then the eigenenergies of the box with the Neumann boundary conditions are given by
\begin{equation}
E_{m,n}=\frac{1}{a_0^2}\left[4-2\cos\left(\frac{\pi(n-1)}{N}\right)-2\cos\left(\frac{\pi(m-1)}{M}\right)\right],
\end{equation}
while the eigen-functions $\psi_{n,m}(j,l)$ are the products of two factors each given by Eq. (\ref{boxeig})
\begin{equation}\label{eig2disc}
\psi_{n,m}(j,l)=\psi_n(j)\psi_m(l)
\end{equation}

After evaluating expressions analogous to Eqs. (\ref{projbox}) and (\ref{projW}) one finds that the coupling between $m,p$ mode of the box and $q$
channel in the left waveguide is
accounted for through the matrix $W_{L,p}(n,m)$
\begin{equation}
W_{L,p}(m,n)=\frac{\psi_n(0)}{\sqrt{a_0}}\sum_l\psi_m(l)\phi_p(l).
\end{equation}
while for the right  waveguide we have
\begin{equation}
W_{R,p}(m,n)=\frac{\psi_n(N-1)}{\sqrt{a_0}}\sum_l\psi_m(l)\phi_p(l).
\end{equation}
The effective non-Hermitian Hamiltonian reads
\begin{equation}\label{discrete}
\widehat{\mathcal{H}}_{eff}=E_{n,m}\delta^{n}_{n'}\delta^{m}_{m'}+\sum_{p=1}^P\frac{(1-e^{ik_pa_0})}{a_0}\sum_{C=L,R}W_{C,p}W_{C,p}^{\dagger}.
\end{equation}
The equation for the interior wave-function $|\chi\rangle$ is
\begin{equation}\label{chi2}
(\widehat{\mathcal{H}}_{eff}-E)|\chi\rangle=-i\sum_{p=1}^{P}\sqrt{\frac{\sin(k_pa_0)}{a_0\pi}}\sum_{C=L,R}W_{C,p}a^{(+)}_{C,p},
\end{equation}
and, finally, for reflection amplitudes $a^{(-)}_{C,p}$ we have
\begin{equation}\label{amp2}
a^{(-)}_{C,p}=-a^{(+)}_{C,p}+\sqrt{\frac{4 \pi\sin(k_pa_0)}{a_0}}W^{\dagger}_{C,p}|\chi\rangle.
\end{equation}

\subsection{Continuous limit}

In the continuous limit $a_0\rightarrow 0$  the eigenfunctions of the resonator $${\psi}_{n,m}(x,y)={\psi}_n(x){\psi}_m(y)$$
are given by
\begin{eqnarray}\label{eig2dcont}
&{\psi}_1(x)=\sqrt{\frac{1}{L_x}}, ~~{\psi}_n(x)=\sqrt{\frac{2}{L_x}}
\cos\left[\frac{\pi n x}{L_x} \right], ~~n>1, x\in [0~ L_x], &\nonumber\\
&{\psi}_1(y)=\sqrt{\frac{1}{L_y}}, ~~{\psi}_m(y)=\sqrt{\frac{2}{L_y}}
\cos\left[\frac{\pi m(y+L_y/2)}{L_y} \right], ~~m>1, y\in [-L_y/2~ L_y/2],
\end{eqnarray}
where we assumed that the $x$-axis coincides with the center-line of the waveguides.
The corresponding eigenvalues are
\begin{equation}
E_{n,m}={\left(\frac{\pi n}{L_x}\right)}^2+{\left(\frac{\pi m}{L_y}\right)}^2.
\end{equation}

The the solutions in the waveguides are
\begin{equation}\label{wavecont1}
\psi_{L,p}^{(\pm)}(x,y)=\frac{1}{\sqrt{4\pi k_p}}e^{\pm ik_px }\phi_p(y), \ x<0,
\end{equation}
and
\begin{equation}\label{wavecont2}
\psi_{R,p}^{(\pm)}(x,y)=\frac{1}{\sqrt{4\pi k_p}}e^{\mp ik_px \pm ik_pL_x}\phi_p(l), \ x>L_x,
\end{equation}
where $\phi_q(y)$ is the transversal wavefunction of the waveguide
\begin{equation}
\phi_1(y)=\sqrt{\frac{1}{d}}, \ \phi_p(y)=\sqrt{\frac{2}{d}}\cos\left[\frac{\pi p(y+d/2)}{d}\right], \ p=2, \ 3, \ \ldots Q.
\end{equation}
The corresponding dispersion relation reads
\begin{equation}
E_p=k_p^2+{\left(\frac{\pi p}{d}\right)}^2.
\end{equation}

Taking into account that the step-size $a_0$ is found as
\begin{equation}
a_0=\frac{d}{P}=\frac{L_y}{M}=\frac{L_x}{N}
\end{equation}
the continuous limit from Eqs. (\ref{wavedisc1}), (\ref{wavedisc2}), and (\ref{eig2disc}) to Eqs. (\ref{eig2dcont}),
(\ref{wavecont1}), and (\ref{wavecont2})  is reached as
\begin{equation}
\psi_{C,p}^{(\pm)}(x,y)=\lim_{a_0\rightarrow 0}\left(\frac{1}{\sqrt{a_0}}\psi_{C,p}^{(\pm)}(j,l)\right), \
{\psi}_{n,m}(x,y)=\lim_{{a_0}\rightarrow 0}\left(\frac{1}{\sqrt{a_0}}{\psi}_{n,m}(j,l)\right),
\end{equation}
Thus, the continuous limit of coupling matrices $\widetilde{W}_{C,q}(m,p)$ could be found as
\begin{equation}\label{WLtilde}
\widetilde{W}_{L,p}(n,m)=\psi_n(0)\int\limits_{-d/2}^{d/2}dy\psi_m(y)\phi_p(y)
\end{equation}
while for the right  waveguide we have
\begin{equation}\label{WRtilde}
\widetilde{W}_{R,p}(n,m)=\psi_n(L_x)\int\limits_{-d/2}^{d/2}dy\psi_m(y)\phi_p(y),
\end{equation}
then the effective non-Hermitian Hamiltonian reads
\begin{equation}\label{Heffcont}
\widehat{\mathcal{H}}_{eff}=E_{n,m}\delta^{n}_{n'}\delta^{m}_{m'}-\sum_{p=1}^{\infty}ik_p\sum_{C=L,R}\widetilde{W}_{C,p}\widetilde{W}_{C,p}^{\dagger}.
\end{equation}
The equation for the scattering function takes the following form
\begin{equation}\label{continuous}
(\widehat{\mathcal{H}}_{eff}-E)|\chi\rangle=-i\sum_{p=1}^{\infty}\sqrt{\frac{k_p}{\pi}}\sum_{C=L,R}\widetilde{W}_{C,p}a^{(+)}_{C,p},
\end{equation}
while the reflection amplitudes are given by
\begin{equation}\label{ampcont}
a^{(-)}_{C,p}=-a^{(+)}_{C,p}+\sqrt{4\pi k_p}~\widetilde{W}^{\dagger}_{C,p}|\chi\rangle
\end{equation}
The interior wave function $\psi_b(x,y)$ could be found as
\begin{equation}\label{solution2d}
\psi_b(x,y)=\sum\limits_{n,m=1}\chi_{n,m}\psi_{m,n}(x,y)
\end{equation}

The above presented approach is valid not
only for the system shown in Fig. \ref{fig3}. In general the resonator eigenfunctions could not be found through
separation of variables as in Eq. (\ref{wavedisc2}). However, on any zero curvature waveguide-resonator interface of length $d$
one can introduce a local set of coordinates with the y-axis directed along the interface while the x-axis is
directed along the waveguide. Then the same calculations give us the the following expression for the coupling matrix
\begin{equation}
\widetilde{W}_{p}(m)=\int\limits_{-d/2}^{d/2}dy\psi_m(y,x)\phi_p(y),
\end{equation}
where we have a single index $m$ to enumerate the eigenfunctions of the resonator. The rest of the formulas (\ref{continuous},
\ref{ampcont},\ref{solution2d}) remain correct with the only difference that index $C$ could run over arbitrary number of
waveguides attached to the resonator. This resulting expressions are consistent with the earlier findings in ref. \cite{Pichugin}.

\begin{figure}[h]
\includegraphics[height=8cm,width=8cm,clip=]{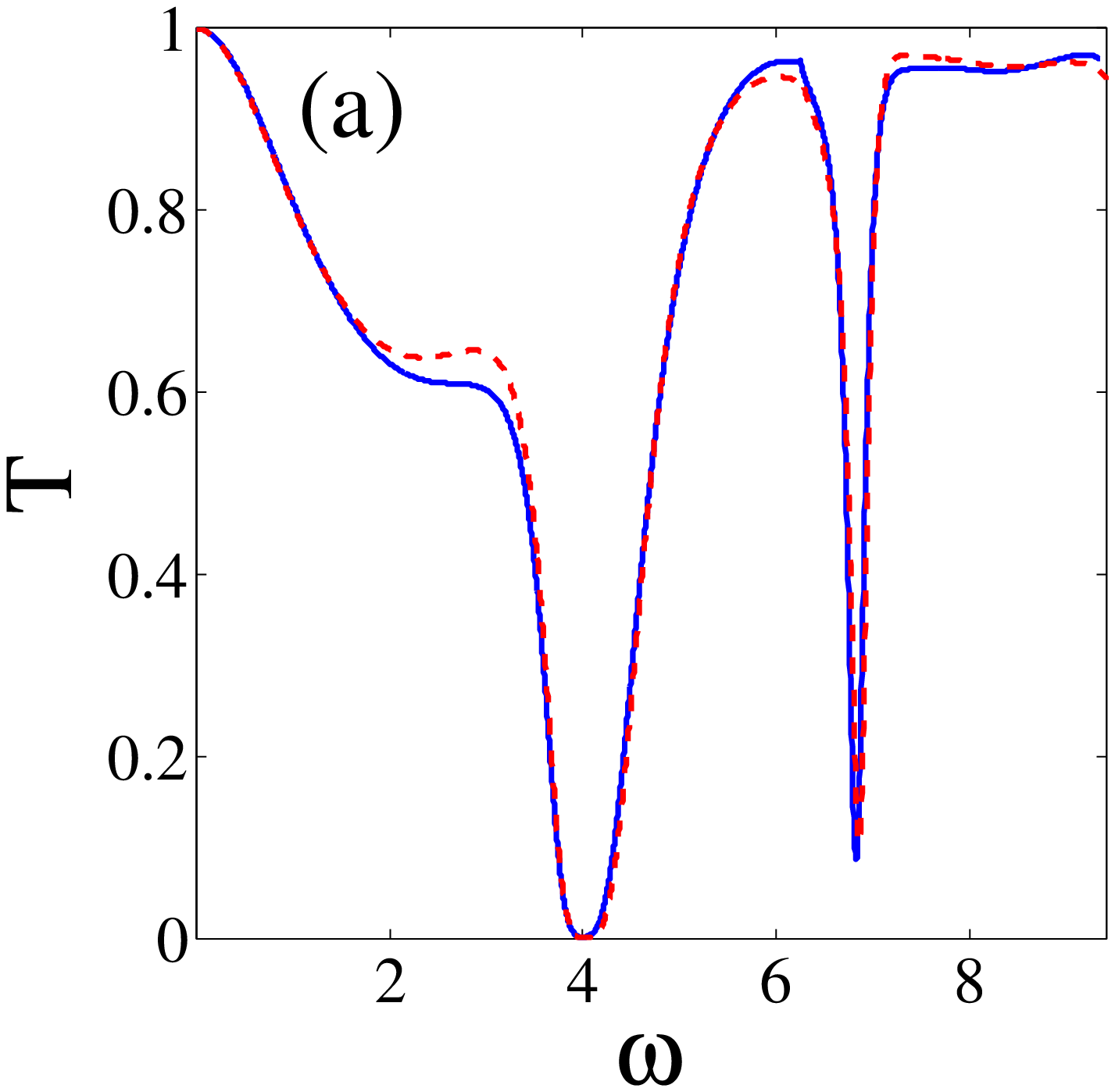}
\includegraphics[height=8cm,width=8cm,clip=]{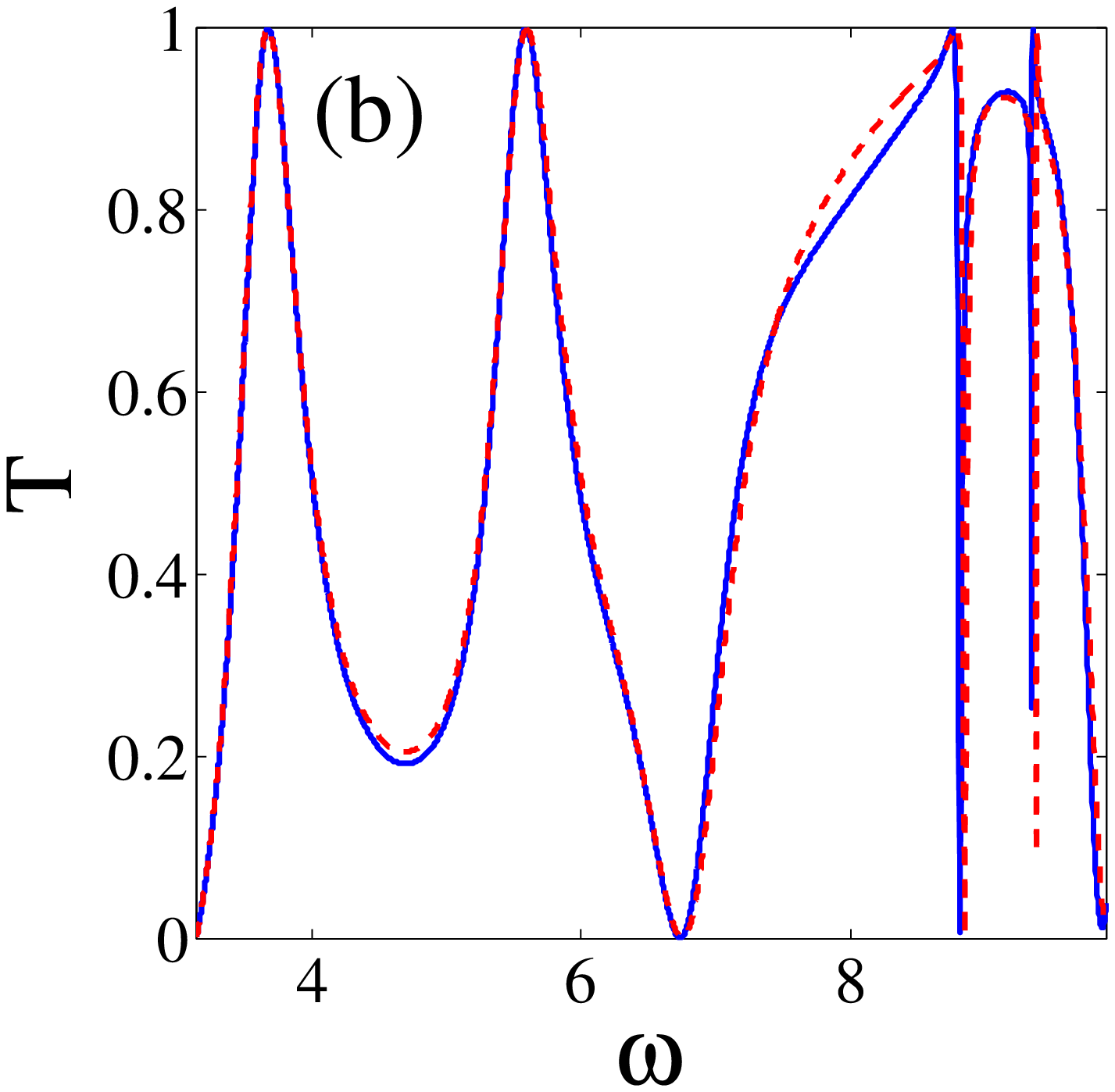}
\caption{(Color online) Transmission spectra of 2D acoustic resonator with the incoming wave in the first (a) and second (b) channels;
discrete model - solid blue line, continuous model - dashed red line.}\label{fig4}
\end{figure}
\begin{figure}[h]
\includegraphics[height=4cm,width=7cm,clip=]{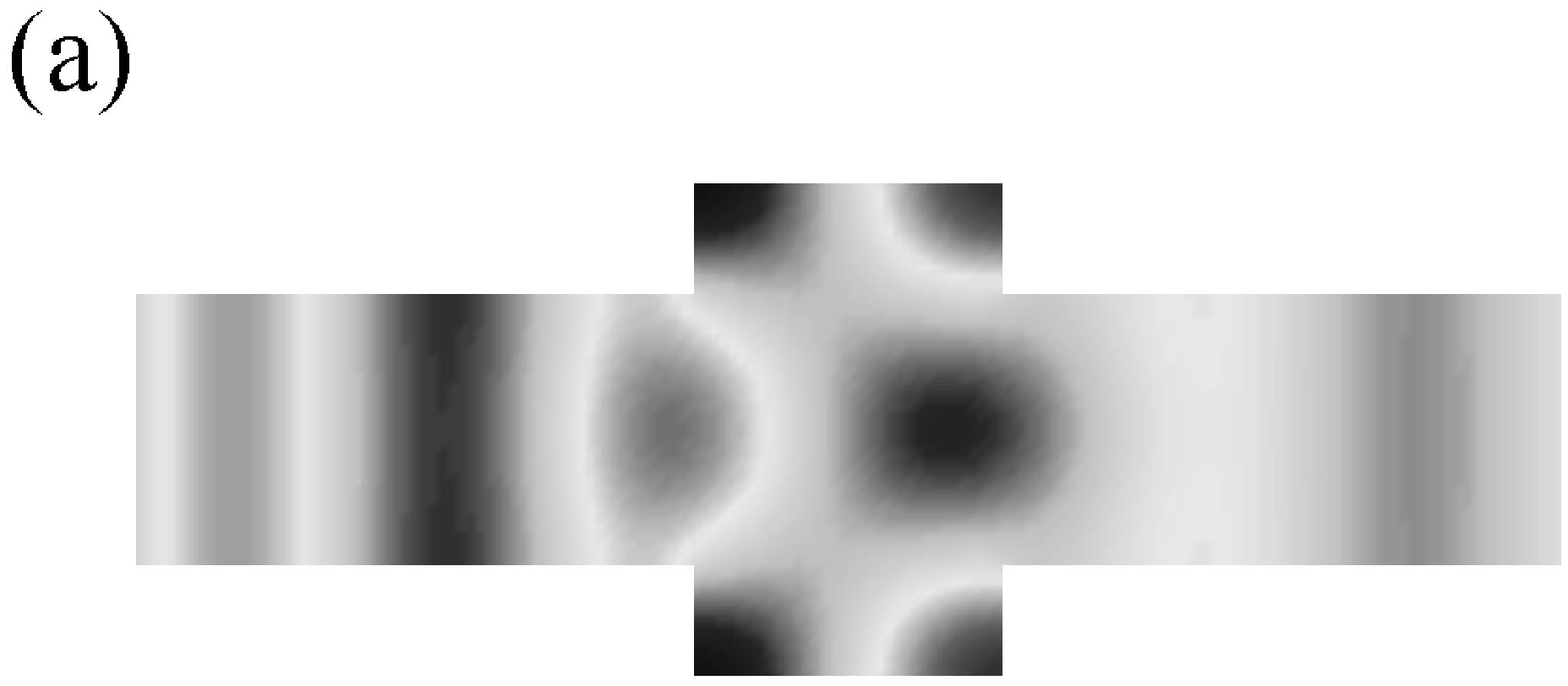}
\includegraphics[height=4cm,width=7cm,clip=]{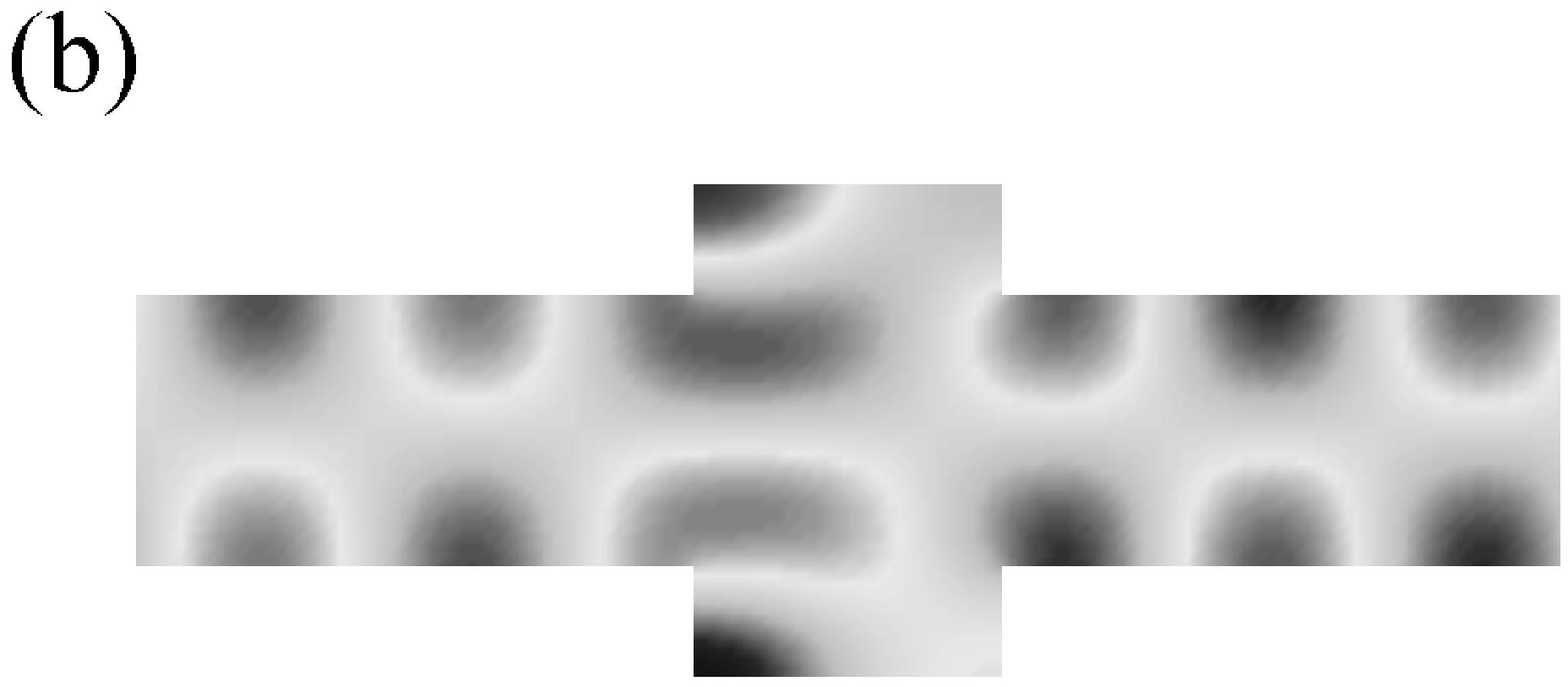}
\caption{Scattering function at $\omega=3\pi/2$ with the incoming wave in the first (a)
and second (b) channels.}\label{fig5}
\end{figure}

\subsection{Numerical results}

To perform a numerical test we calculated the transmission coefficient $T$ using both discrete (\ref{discrete}) and continuous (\ref{continuous}) approaches.
The geometric parameters of the system were chosen as $d=1, L_y=1.8, L_x=1$. In the discrete model we took $N=20, Q=20, P=36$ which corresponds to $a_0=0.05$.
In the continuous model where we have an infinite number of modes in the resonator as well as in the waveguide Eq. (\ref{continuous}) was truncated
to a finite number of unknowns. To be consistent with our discrete model we took $20$ modes in each waveguide while the indices $m$ and $p$ in Eq. (\ref{continuous})
run as $m=1, 2, \ldots, M$ and $n=1, 2, \ldots, N$. Thus, we guaranteed that in both approaches we have the same number of degrees of freedom. For our system
the coupling matrices $\widetilde{W}_{L,p}(n,m)$ and $\widetilde{W}_{R,p}(n,m)$ could be found analytically through Eqs.
(\ref{WLtilde}) and (\ref{WRtilde}). The
resulting expressions are
\begin{equation}
\widetilde{W}_{L,p}(n,m)=\sqrt{\frac{2-\delta_n^{0}}{L_x}}V_{p,m}; \ \widetilde{W}_{L,p}(n,m)=\sqrt{\frac{2-\delta_n^{0}}{L_x}}{(-1)}^{n-1}V_{p,m},
\end{equation}
where
\begin{equation}
V_{p,m}=\left\{\begin{array}{ccc}
\sqrt{\frac{d}{L_y}} & \mbox{if $p=1, m=1$};
\\ 0 & \mbox{if $p>1, m=1$};
\\  \frac{1}{\pi m}\sqrt{\frac{2L_y}{d}} \left(\sin\left[\pi m \frac{L_y+d}{2L_y}\right]-\sin\left[\pi m \frac{L_y-d}{2L_x}\right] \right)& \mbox{if $p=1, m>1$};
\\ \frac{2\sqrt{L_yd}dm}{\pi\left({(md)}^2-{((p-1)L_y)}^2 \right)} \left((-1)^{p-1}\sin\left[\pi m \frac{L_y+d}{2L_y}\right]-\sin\left[\pi m \frac{L_y-d}{2L_y}\right] \right)& \mbox{if $p>1, m>1$}.
\end{array}\right.
\end{equation}
The numerical data are shown
in Fig. \ref{fig4} where one can see a very good agreement between two methods. This, in accordance with ref. \cite{Pichugin}, proves that the effective non-Hermitian Hamiltonian
is stable with respect to truncation of the number of modes. Finally, in Fig. \ref{fig5} we present two typical configurations of the real part of the pressure field
with the resonator excited through both first and second scattering channels.

\section{Transmission through cylindrical acoustic resonator} \label{sec4}
\begin{figure}[ht]
\includegraphics[height=6cm,width=6cm,clip=]{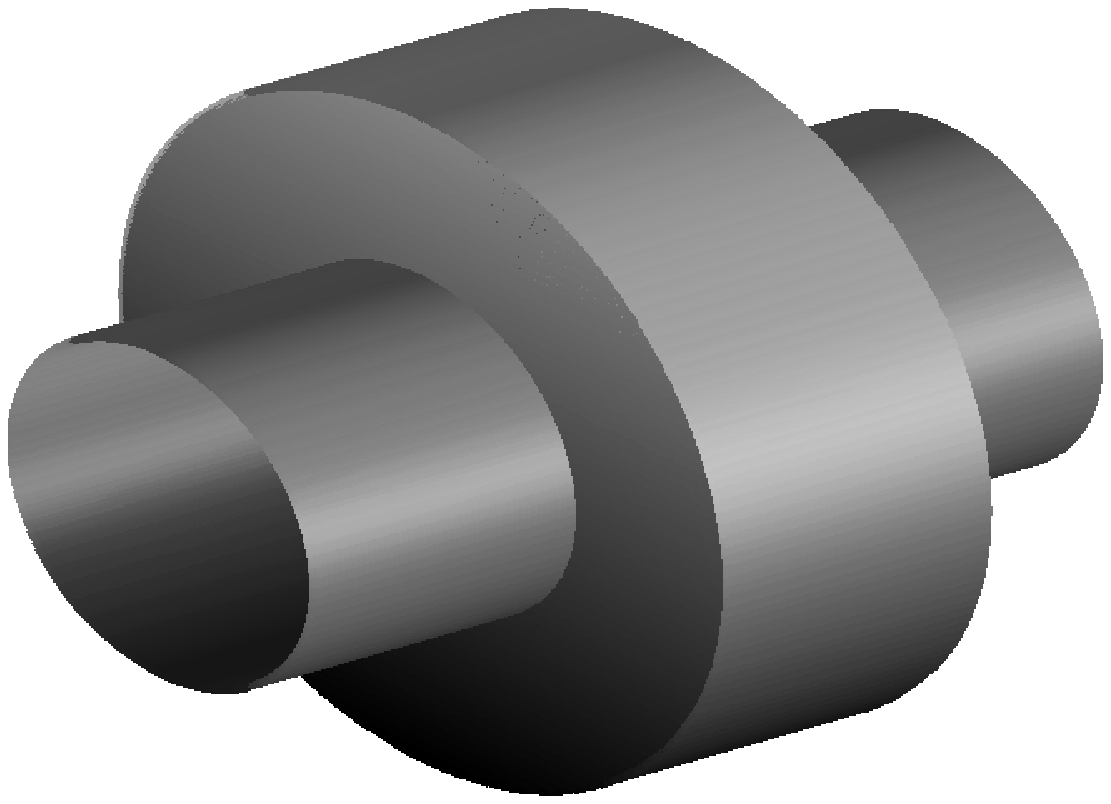}
\caption{Cylindric resonator of radius $R$ and length $L$ connected to coaxial waveguides with radius $a$.}\label{fig6}
\end{figure}

In this section we consider sound transmission in three dimensional system, namely
a cylindrical resonator coupled with two cylindrical waveguides as shown in Fig. \ref{fig6}.
We restrict ourself to the coaxial case where the angular momentum
$-i\frac{\partial}{\partial \phi}$ is a constant of motion $\phi$ being the azimuthal variable. Therefore the Hilbert space of
the total system is a direct sum of the subspaces specified by the angular momentum.
Below we present the effective non-Hermitian Hamiltonian approach in case of zero angular momentum.
The generalization to non-zero angular momentum is straightforward.

\subsection{The effective non-Hermitian Hamiltonian}

The eigenfrequencies of the closed cylindrical resonator with the Neuman boundary conditions
are given by
\begin{equation}\label{eigcyl}
\omega_{mn}^2=c^2\left[\frac{\mu_{m}^2}{R^2}+\frac{\pi n^2}{L^2}\right]
\end{equation}
where $R$ and $L_z$ are the radius and length of the cylindrical resonator, and $\mu_m$ is the m-th root of the
equation $J_0'(\mu_m)=J_1(\mu_m)=0, m=1, 2, \ldots$ for the derivative of the first order Bessel function. The corresponding eigenfunctions are
\begin{equation}\label{eigfun}
    \psi_{mn}(r,\phi,z)=\frac{1}{\sqrt{\pi}RJ_0(\mu_m)}J_0\left(\frac{\mu_mr}{R}\right)\psi_n(z), \psi_1=\sqrt{\frac{1}{L_z}},
    \psi_n(z)=\sqrt{\frac{2}{L_z}}\cos\left(\frac{\pi (n-1) z}{L_z}\right), n=2, 3, \ldots.
\end{equation}
Similar to the 2D acoustic case considered in the previous section in the continuous limit
we obtain the effective Hamiltonian in the same form as Eq. (\ref{Heffcont})
with the coupling matrices given by
\begin{equation}
\widetilde{W}_{L,p}(n,m)=\sqrt{\frac{2-\delta_n^{0}}{L}}V_{p,m}; \ \widetilde{W}_{L,p}(n,m)=\sqrt{\frac{2-\delta_n^{0}}{L}}{(-1)}^{n-1}V_{p,m},
\end{equation}
with
\begin{equation}\label{Wcyl}
    V_{p,m}=\frac{1}{\pi aR J_0(\mu_m) J_0(\mu_p)}\int\limits_0^a\int\limits_0^{2\pi}drd\phi rJ_0(\mu_mr/R)J_0(\mu_pr/a),
\end{equation}
where $a$ is the radius of the cylindrical waveguides.
Integration in Eq. (\ref{Wcyl}) can be done analitically \cite{Prudnikov}
\begin{equation}\label{Bessels}
    V_{p,m}=\frac{2a^2\mu_m}{{(\mu_ma)}^2-{(\mu_pR)}^2}\frac{J_1(\mu_ma/R)}{J_0(\mu_m)}.
\end{equation}
Finally, the dispersion relation for the scattering channels reads
\begin{equation}\label{contpipe}
\omega^2=c^2(\mu_{p}^2/a^2+k_p^2), p=1,2,\ldots  .
\end{equation}
The numerical procedure for finding the reflection amplitudes $a_{p,c}^{(-)}$ and the interior scattering
function is identical to the 2D case considered in the previous section Eqs. (\ref{Heffcont}), (\ref{continuous},
(\ref{ampcont}), and (\ref{solution2d})

\subsection{Mixed representation}
\begin{figure}[ht]
\includegraphics[height=8cm,width=8cm,clip=]{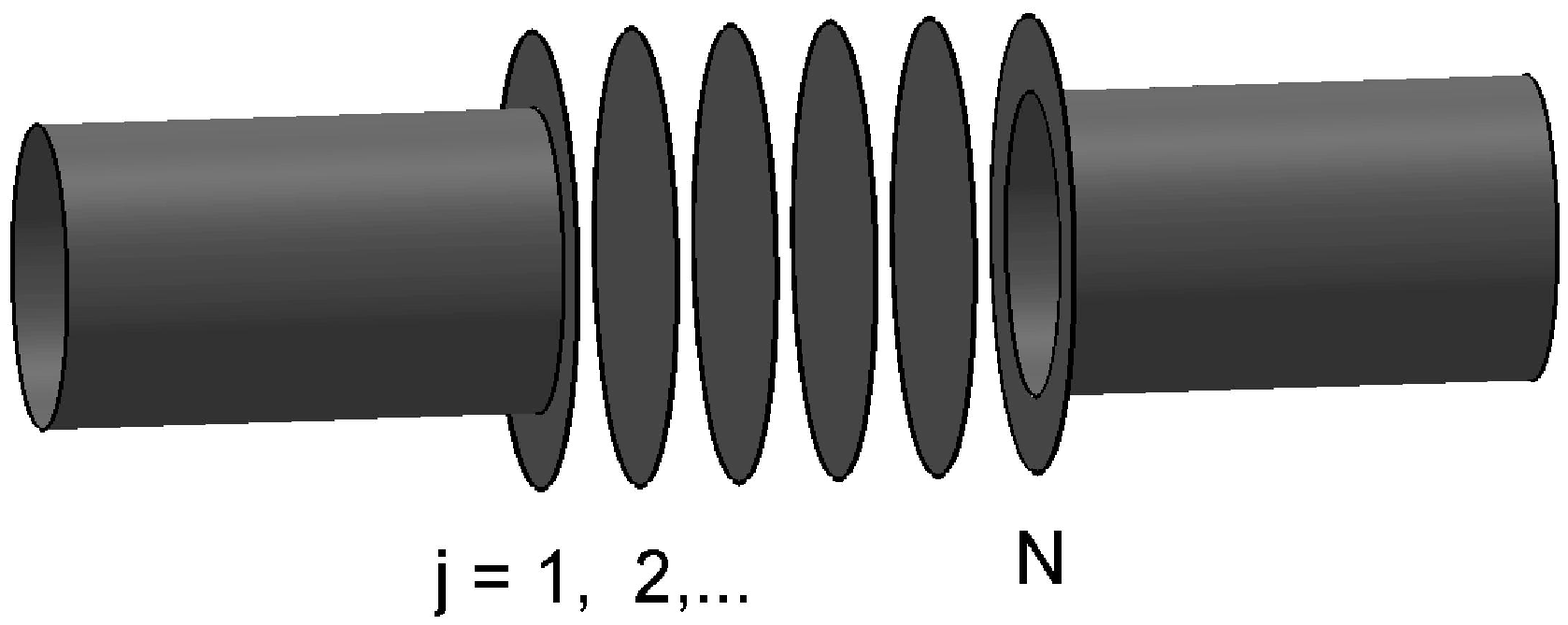}
\caption{Cylindric resonator discretized in the axial coordinate.}\label{fig7}
\end{figure}
Another interesting opportunity for separable systems is to use the discretized representation in one coordinate
while the continuous representation is used in the other coordinates.
Let us write the Helmholtz equation (\ref{Helmholtz}) in the following form
\begin{equation}\label{WE}
\frac{1}{a_0^2}\left(\left[\frac{\partial^2}{\partial r^2}+\frac{1}{r}\frac{\partial}{\partial r}\right]\psi_j
-2\psi_j+\psi_{j-1}+\psi_{j+1}\right)+E\psi_j=0,
\end{equation}
where we used the continuous representation for the radial and azimuthal coordinates and the discrete representation for
the axial coordinate $z$. The idea of the approach is sketched in Fig. \ref{fig7}.
The resulting expressions are the same as those found in \ref{discretesub}  for both scattering function Eq. (\ref{chi2}) and
reflection amplitudes Eq. (\ref{amp2}) with the only difference that the coupling matrices are now given by
\begin{equation}
W_{L,p}(m,n)=\frac{\psi_n(0)}{\sqrt{a_0}}V_{p,m}, \
W_{R,p}(m,n)=\frac{\psi_n(N-1)}{\sqrt{a_0}}V_{p,m},
\end{equation}
where $V_{p,m}$ are defined through Eq. (\ref{Bessels}).
The eigenfrequincies of the resonator are now found as
\begin{equation}
\label{eigdisc}
\omega_{mn}^2=c^2\left[\frac{\mu_{m}^2}{R^2}+\frac{2-2\cos(k_na_0)}{a_0^2}\right],
k_n=\frac{\pi (n-1)}{L}, n=1,2,\ldots, N,
\end{equation}
while the dispersion relation for the channel function is
\begin{equation}\label{contdisk}
\omega^2=c^2\left[\frac{\mu_{p}^2}{a^2}+\frac{2-2\cos(k_pa_0)}{a_0^2}\right], p=1, 2, \ldots.
\end{equation}

\subsection{Numerical results}
\begin{figure}[h]
\includegraphics[width=8cm,clip=]{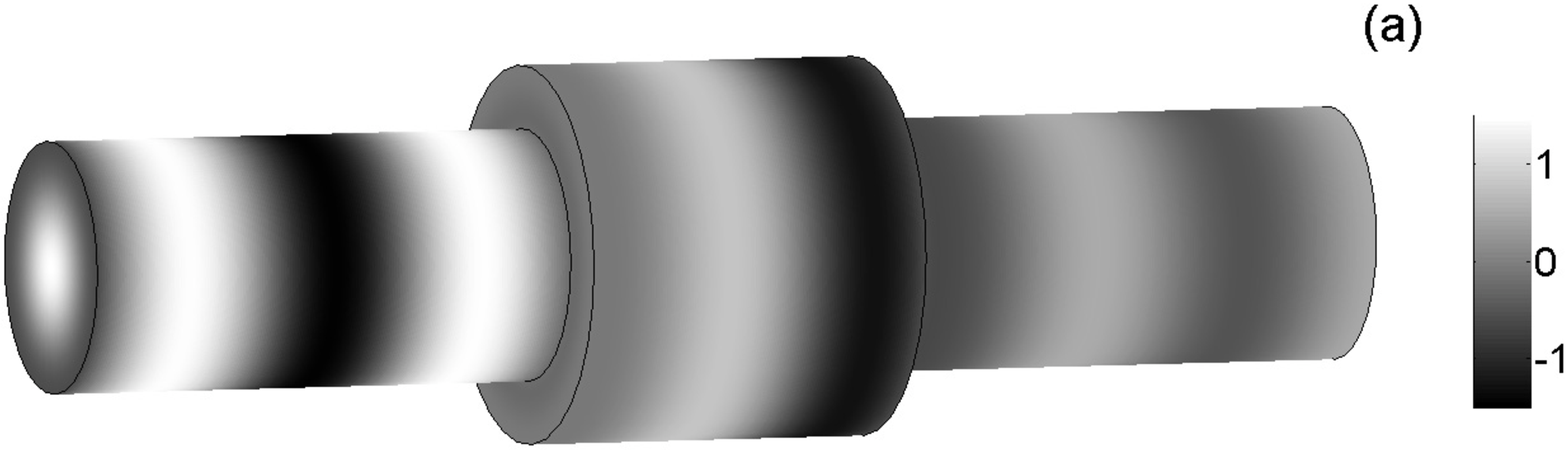}
\includegraphics[width=8cm,clip=]{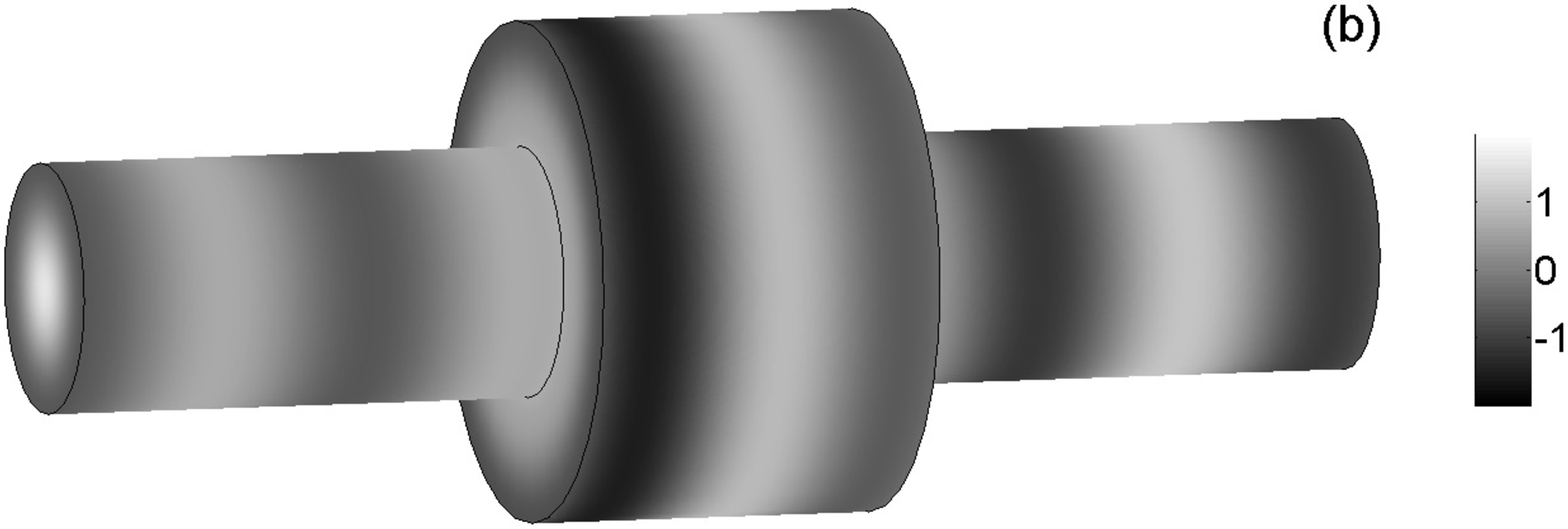}
\caption{Scattering function (real part)
in a cylindrical resonator coupled with two coaxial waveguides. The wave is incident through the fist channel.
$L=3.5, \omega=6.6845$. (a) $R=1.5a$ and (b) for $R=2a$.} \label{fig8}
\end{figure}
\begin{figure}[ht]
\includegraphics[width=8cm,clip=]{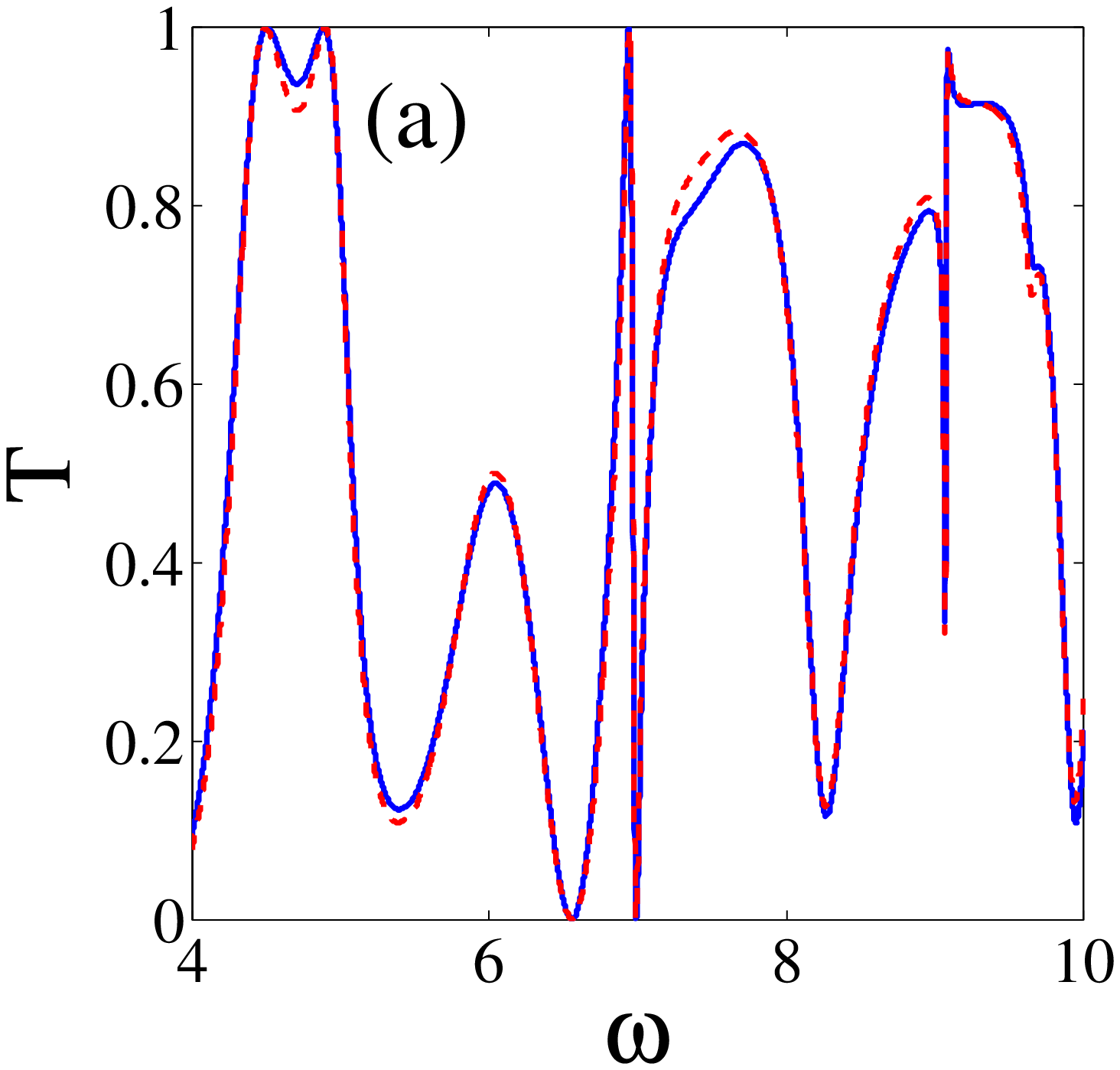}
\includegraphics[width=8cm,clip=]{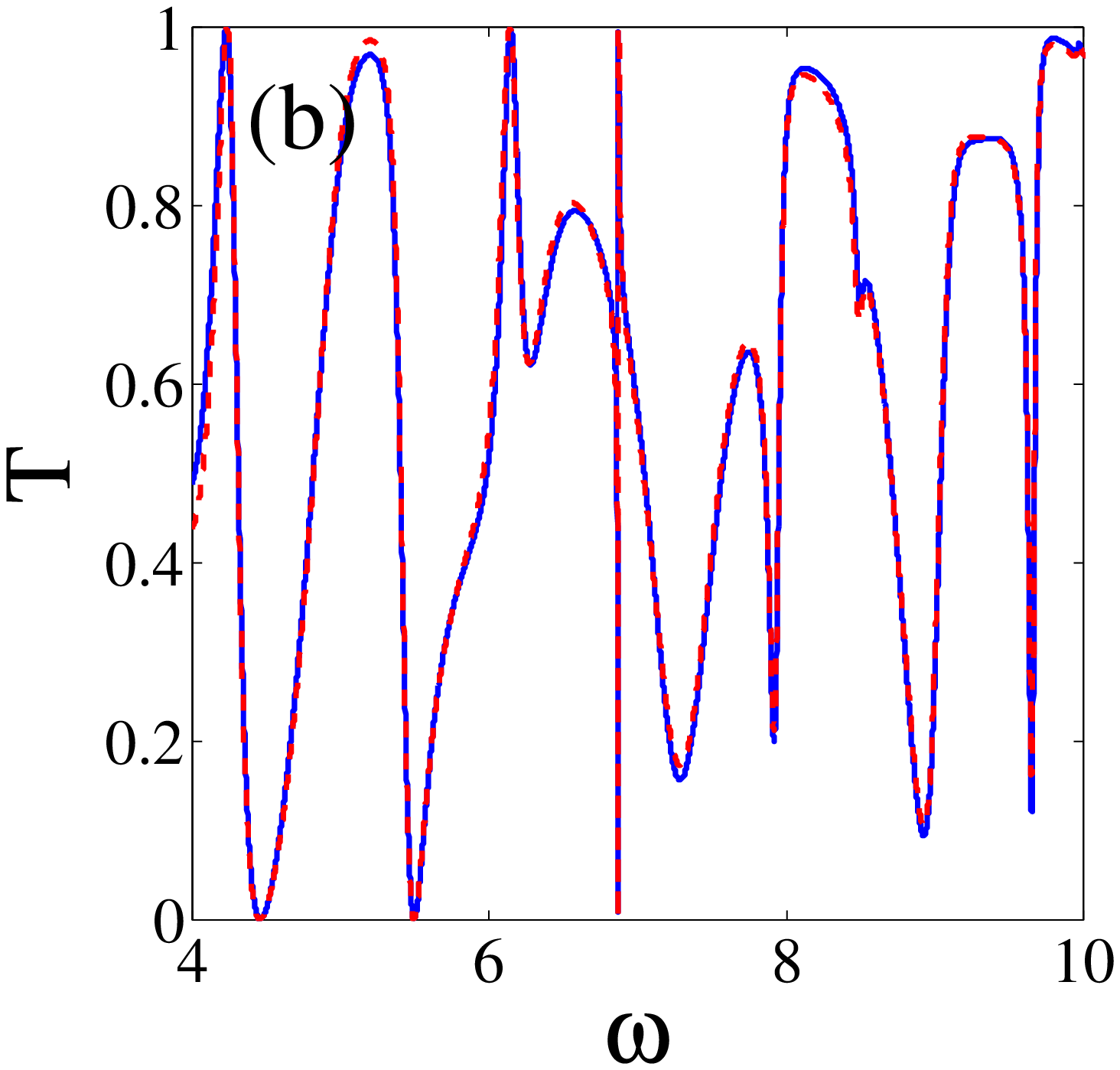}
\caption{(Color online) Transmission spectra of cylindric acoustic resonator with two differing aspect ratios:
(a) $R/a=1.5$ and (b) $R/a=2$. Continuous representation - solid blue line, mixed representation - dashed red line.}\label{fig9}
\end{figure}

Numerically computed scattering function are shown
in Fig. \ref{fig8}. Here, to perform the numerical test we took $10$ modes in each waveguide while in the resonator
$20$ modes were taken in each radial and axial variables. The same numbers were chosen in the mixed representation.
In Fig. \ref{fig9} we demonstrate the transmission spectra of the cylindrical resonators
with two differing aspect ratios $a/R$.  One can see a
good agreement between continual and mixed approaches even for relatively small number of slices $N=20$ (see Fig. \ref{fig7}).

\section{Summary} \label{sec5}
In this paper we developed the approach of the effective non-Hermitian Hamiltonian to resonator-waveguide systems
with the Neumann boundary conditions. To derive the effective non-Hermitian Hamiltonian we used an alternative approach
which dates back to the seminal paper by Fano \cite{Fano} and was recently adapted for the discrete Schr\"odinger equation \cite{Maksimov}.
In contrast with the standart Feshbash projection technique \cite{Feshbach} our approach does not face a problem of principal value
integration. It was shown that the method of the effective non-Hermitian Hamiltonian can be used to calculate the $S$-matrix
and the scattering function of sound-hard boundary acoustic scattering problem.
Thanks to the projection of the total Hilbert space onto the inner states of the resonator the method is free of the discontinuity problem typical
for the mode-matching techniques \cite{Wexler,Bornemann,Muehleisen,Solokhin,Sharma,Buyukaksoy,Kirby,Snakowska}.
It was shown that in the continuous case the effective Hamiltonian can be truncated to
a reasonable small number of eigenmodes of the closed system whose eigenfrequencies are concentrated around
the frequency of the incident wave. In that sense the method of effective non-Hermitian Hamiltonian is analogous to
the coupled-mode theory \cite{Haus,Suh} which however neglects the radiation shifts caused  by a
finite lower band edge and dispersive properties of the waveguides. Therefore we can refer to the approach of the effective non-Hermitian Hamiltonian
as an advanced form of the coupled mode theory. In this paper the effective non-Hermitian Hamiltonian was
formulated in both continuous and discrete forms. The former is shown to be consistent with the earlier findings
in ref. \cite{Pichugin} while the latter seems to be more feasible for numerical implementation along
with finite-difference and finite-element discretization schemes. In fact, the developed approach represents
a tool for modelling systems with open boundaries. Alike to the recently developed wave finite-element methods \cite{Waki, Renno}
it explicitly utilizes the translation invariance in the waveguide to formulate numerically exact transparent
boundary conditions \cite{Ando}. It should be noted, however, that the resulting equations (\ref{LS2}), (\ref{chi2}),  and
(\ref{continuous}) contain dense rather then sparse matrices typical for the lattice methods. This to a certain degree degrades the numerical performance.
Nevertheless, thanks to a clear physical picture of the wave transmission one may think of developing techniques based
on truncation of the inner space to a small number of modes which are relevant to a particular case of the scattering problem.
This represents a goal for the future studies. Finally, the effective Hamiltonian has an advantage in physical
interpretation of it complex eigenvalues $z$. The real parts $E_r=Re(z)$  define the positions of resonances with the
resonance widths defined by $-2Im(z)$ \cite{Ingrid}. That advantage becomes important in
study of wave trapping \cite{RS,Evans,Hein} by tracing the imaginary parts of complex
eigenvalues of the effective non-Hermitian Hamiltonian.

 {\bf Acknowledgments}. We thank K.N. Pichugin  for helpful discussions.
 The work was supported by grant 14-12-00266 from Russian Science Foundation.

\bibliographystyle{elsarticle-num}
\bibliography{wave_motion}

\end{document}